\documentclass[doublespacing]{elsart}

\usepackage{natbib, graphicx}

\usepackage{amssymb}
\journal{Physica A}
\begin{document}

\begin{frontmatter}

\title{Time-evolving distribution of time lags  between commercial airline disasters}

\author[lambi]{M. Ausloos},
\corauth[cor]{Corresponding author.} 
\ead{marcel.ausloos@ulg.ac.be}
\author[lambi]{R. Lambiotte},
\ead{Renaud.Lambiotte@ulg.ac.be}

\address[lambi]{SUPRATECS,  B5 Sart-Tilman, B-4000 Li\`ege, Euroland}

\begin{abstract}
  We have studied the time lags between commercial line airplane disasters and their occurrence frequency till 2002, as obtained from a freely available website.
  We show that the time lags seem to be well described by Poisson random events, where the average events rate is itself a function of time, i.e. time-dependent Poisson events.  This is likely due to the unsteady growth of the industry. The time lag distribution is compared with a truncated 
   Tsallis distribution, thereby showing  that the ''phenomenon'' has similarities with  a Brownian particle with time dependent mass. We distinguish between ''other causes'' (or natural causes) and ''terrorism acts", the latter amounts to about 5 percents,
   but we find no drastic difference nor impact due to the latter on the overall distribution.
\end{abstract}

\begin{keyword}
Poisson distribution \sep exponential distribution \sep Tsallis distribution \sep Airline flight crashes 
\sep Terrorism
\PACS 45.70.Vn\sep 05.40.Fb\sep 05.65.+b

\end{keyword}

\end{frontmatter}

\section{Introduction}
One modern question in statistical mechanics pertains to the extreme events and subsequent risk of avoidance. 
The distribution of  events is described by various event probability distribution functions (PDF). 
No need to recall that the tails of the PDF are carefully examined in many fields of science, politics, psychology, economy ...
For very large numbers of observations, due to the central limit theorem, the Gaussian law is the theoretically expected one. 
The log normal distribution is very similar to the Gaussian\cite{gregory}, and is better used when experiments can be often repeated.  

The Poisson distribution was introduced for describing the number of deaths $n$ during a given time interval by horse kicking (in the Prussian army) \cite{gregory}, 
and is well-known to prevail for independent and rare events \cite{poissondistrwebsite}. In general, it reads

\begin{equation}
 P(n|a)=\frac{a^n}{n!} e^{-a},
 \end{equation}
where $n$ is the number of events occurring during some time interval, and $a$ is the arithmetic average of $n$.
It is well-known that the waiting times between two successive Poisson events distribute like a negative exponential:
\begin{equation}
f(\tau)= \tau_{c}^{-1} \exp{(- \tau/ \tau_{c}).}
\end{equation}
where $\tau_C$ is the average characteristic waiting time between events.
Amongst others, these statistics have been used for describing nuclear desintegration, 
i.e.  the frequency of nuclear events occurred in time intervals given by the Poisson function, but also for the time lags between shoppers entering a store, 
the number of phone calls in a time interval, the number of failure of products in a time interval,  
and also for spatial interval distributions like the fall of meteorites on land \cite{gregory}.

Do plane  accidents or more specifically plane crashes  enter this category of independent critical events? 
It seems that this simple question has not been much studied up to now, thereby requiring the characterisation 
of the time separation between (or frequency of) such crashes. Yet recent events seem to suggest such an investigation, 
i.e. recall the case of 5 major crashes in August 2005, when this paper was being completed/reviewed.
One may also wonder whether specific external fields influence the distribution of plane crashes. 
Disregarding the time of classical wars, one is aware that so called  terrorists attempted to satisfy some psychological 
and other conditions by putting bombs in planes and exploding them. One may question whether the time distribution 
between plane crashes depends on such acts. The more so at this time of so-called war.

One could argue that plane accident should be separated between those having led to human casualties and others without 
casualties \cite{wiki,NTSB}. The definition of the plane is also relevant: airplane, helicopters, gliders, dirigibles, ULM, .... 
One could demand some data analysis on commercial airplanes, as well as about  chartered, private or military ones. 
These are hard to obtain. The various causes of accidents might be distinguished, - when furnished by the inquiry conclusions. 
Recall for instance the Lauda Air plane explosion over Thailand on May 26, 1991 (223 deaths), first thought to be a terrorist act which targeted the wrong plane/flight, but  the disaster was 4 years later attributed to electromagnetic interference, from a camcorder, laptop computer or mobile phone, with the plane electronic equipment \cite{lauda}. 

In order to perform this time lag analysis, we have examined a freely available data set of airplane disasters, involving commercial passengers,  
from the point of view of time intervals between such events, but also examining whether criminal acts influence the distribution,
 when the voluntarily brutal destruction of the carrier is acknowledged and recognized as such. In so doing we somewhat examine
  endogenous and exogenous cases \cite{RLMAPhysAshocks,sornetteofcourse} of disaster, - the exogeneous cause being here quite specified.  Let it be known at once that this ''sabotage'' (in a broad sense) amounts to only about 0.05 to 0.08 of disasters (between 1950 and 2004), according to $http://www.planecrashinfo.com/cause.htm$ \cite{NTSB}.

\begin{figure}
\includegraphics[angle=-90,width=5.00in]{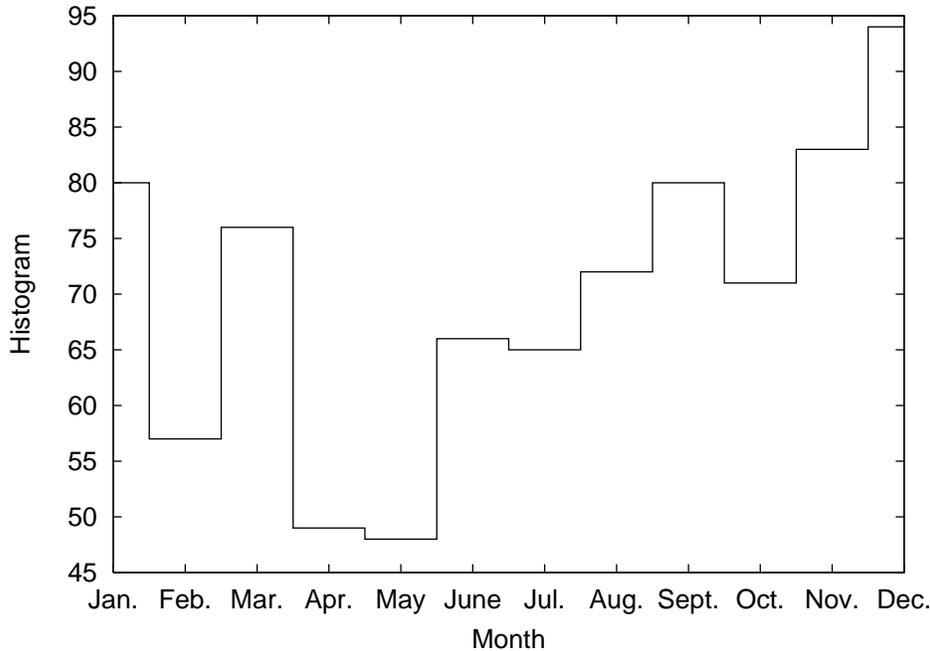}

\caption{\label{month} Histogram of the number of commercial airline disasters as a function of the month when they occurred 
for the [1920-2002] data examined in the text. }
\end{figure} 

$A$ $priori$  one might imagine that plane crashes are independent events. However media discuss the occurrence of series, or avalanches,
- as in Aug. 2005. What are the characteristics of such series? Are they existing?  Do they result from endogenous or exogenous causes,
 or are the series mere illusions? Correlations with cosmic or geophysical activity might be considered. 
 This suggests to examine data 
 through time filters, like done for financial indices, sometimes attempting to forecast financial crashes,
  and correlate the data to other time series. In this first paper, we will not attempt to correlate the data
   on the examined crashes with any natural cause. Apparently there are more disasters during the Northern Hemisphere
    fall-winter time with a high peak occurring in December (Fig. 1). 

An interesting fundamental question is nevertheless of great interest. The usual distribution functions are measured
 for systems in which several quantities are constrained through conservation laws. It is quite clear that the number
  of airplanes and flights have increased through the last century. It is of interest to observe whether such an evolution
   influences the expected Poisson distribution. Considering that the number of planes is not conserved, and a plane is a  ''particle'',
     one might admit that the system is out of equilibrium. Whence characteristic distribution functions for such systems might be considered,
      like the Tsallis one \cite{Tsallisdistribution,tsallis2}; we will do so.

A final warning is necessary, like in many not reproducible (off-laboratory) experiments, the data set completeness,
 accuracy, whence validity can be questioned. We have taken data e-published by {\em http://dnausers.d-n-a.net/dnetGOjg/
Disasters.htm} \cite{datawebsite}.  The data set gives the date of the crash together with some reason for the latter, and the number of deaths,
 beside the type of plane or airline.  We have examined the data in order to remove spurious events during the numerical analysis, e.g. a collision between two airplanes is only considered as one single event. Sometimes the data table mentions $mid$ $air$ $collision$, but mentions only one plane. We cannot guarantee that all crashes have been recorded on this website. It is therefore expected that our study will lead to further investigations which with better data informations might lead to different results and conclusions.

In sect. 2, we analyse the data obtained from a web site.  We show that the airline crash distribution is not Poisson-like, Eq.(1), and equivalently that the time lag distribution exhibits strong deviations from the negative exponential Eq.(2). Due to the small amount of data, we focus on this time lag distribution and introduce a risk function in order to study the features of the distribution tail. By doing so, we find that crash statistics are characterised by a time-dependent exponential distribution function,
  and  seem to be well-described by a Tsallis distribution.
Let us also stress that the mechanisms leading to the anomalous time distribution are very similar to those occurring
 in the case of a Brownian particle with a fluctuating mass \cite{RLMAunpublished}.
In sect. 3, we single out the few disasters considered to have resulted from criminal acts. We find quasi exactly the same characteristics
 as in the ''natural cause'' cases.
Sect. 4 serves as a  brief conclusion.

\section{Data Analysis}

Data for aircraft crashes with announced 150 or more deaths, and a few others, are available between 1921 and 2004 on  {\em http://www.infoplease.com/ipa/A00014
49.html} \cite{siteipa}.
Another web site   {\em http://www.scaruffi.com/politics/aircrash.html} is supposed to contain a list of airplane accidents that caused  50 victims or more between 1946 and 2005 \cite{sitescaruffi}.  A third website  {\em http://dnausers.d-n-a.net/dnetGOjg/
Disasters.htm}  covers the interval time :   Dec. 14, 1920 - March 22, 2002, and seems to be one giving the information on commercial airplanes  only \cite{datawebsite}. 
 See also $http://www.
 planecrashinfo.com/cause.htm$ for a list of 2147 accidents from 1950 thru 2004, from $PlaneCrashInfo.com$ accident database  \cite{NTSB}.
 Surprisingly the lists are not quite identical even when they should be expected to be, either on date or on the number of casualties, - or on causes. The variations appear at first sight to be mild in a statistical sense, whence would likely not  seriously  damage the conclusions of the present analysis.
We have chosen to analyze the data extracted from {\em http://dnausers.d-n-a.net/dnetGOjg/Disasters.htm} over the interval time [Dec. 14, 1920 - March 22, 2002], i.e. an interval of 29 344 days during which there were 
841 crashes \cite{datawebsite}.

\begin{figure}
\includegraphics[angle=-90,width=5.00in]{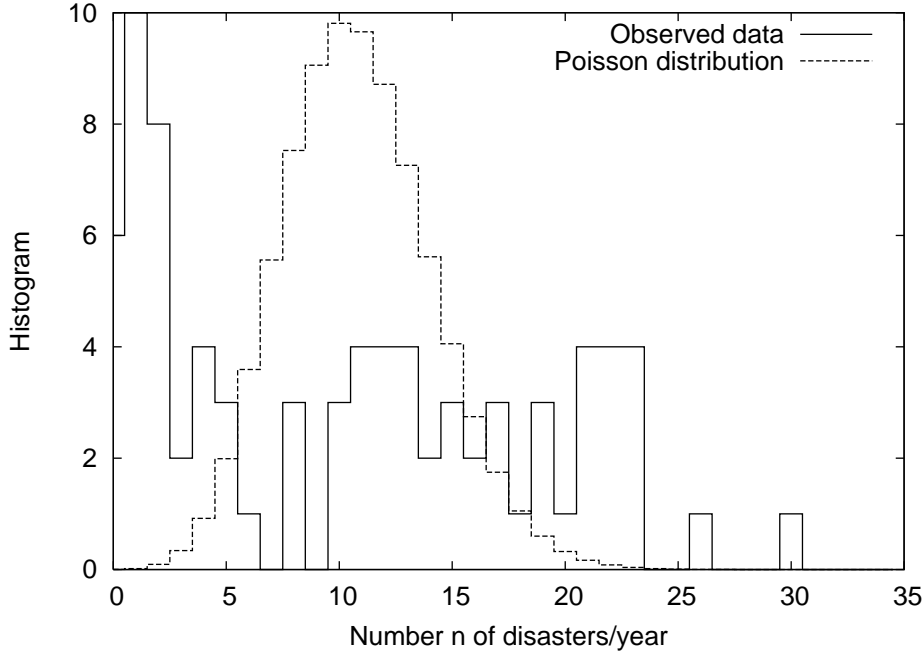}

\caption{\label{poissonG} Number of times that there are $n$ crashes in a given year. We compare the empirical results with an unnormalized Poissonian distribution with the same average $<n>=10.83$ as the empirical data. There are obviously strong deviations between the two distributions. }
\end{figure} 

In order to characterise the time lags $\tau$ between two consecutive crashes, we first focus on 
 the average time lags $ \tau_A=<\tau> \sim 35$ days, so that the average frequency of such events is $\lambda_A=0.0286$ (days$^{-1}$) $\equiv \tau_A^{-1}$. A more complete description of the time intervals requires the study of the following rescaled moments $m_i= \frac{<\tau^i>}{<\tau>^i} - M_i$, where $M_i$ is the value of $\frac{<\tau^i>}{<\tau>^i}$ calculated for the exponential distribution $\tau_A^{-1} \exp(- \tau/\tau_A)$. It is straightforward to show that $M_i = i!$. Let us note that these quantities measure deviations from the theoretical distribution, and have a significance similar to that of the kurtosis and skewness for comparison with Gaussian statistics on the whole real axis $[-\infty, \infty]$. Indeed, positive values of $m_i$ indicate a fat tail of the distribution (as compared to the exponential), while negative values correspond to underpopulated tails.
Our empirical results give the following first moments:
\begin{eqnarray}
m_1&=&0 \cr
 m_2&=&3.12 \cr
 m_3&=&57.56
\end{eqnarray}
where the first moment vanishes by definition. The higher moments imply that the time lag distribution exhibits an overpopulated tail, i.e. emphasizing ''rather long'' times between events.
This suggests that commercial airline crashes are $not$ distributed in a usual Poisson way. This feature is verified by comparing a histogram of the number 
  of times  there are $n$ crashes in a given year, withe ''equivalent'' Poisson distribution, as shown in Fig.\ref{poissonG}. 

Given the lack of available data, let us now consider the probability $f(\tau)$ for {\em waiting times} in more detail, and let us examine the cumulated distribution $P(t)$: 

\begin{equation}
\label{cumulated}
P(t)=\int_0^t f(\tau) d\tau.
\end{equation}
By construction, this function converges toward 1.
In order to study the asymptotic relaxation of Eq.(\ref{cumulated}), it is convenient to  focus on the {risk function} \begin{equation}  G(t)=1-P(t).\end{equation}

Let us note that $G(t)$ converges to zero for $t \rightarrow \infty$ in a related way as the time distribution $f(\tau)$ does:

\begin{eqnarray}
G(t) \sim \exp(-\mu t) &\Leftrightarrow& f(\tau) \sim \exp(-\mu \tau) \cr
G(t) \sim t^{-\alpha} &\Leftrightarrow& f(\tau) \sim \tau^{-(\alpha+1)}.
\end{eqnarray}

\begin{figure}

\includegraphics[angle=-90,width=5.00in]{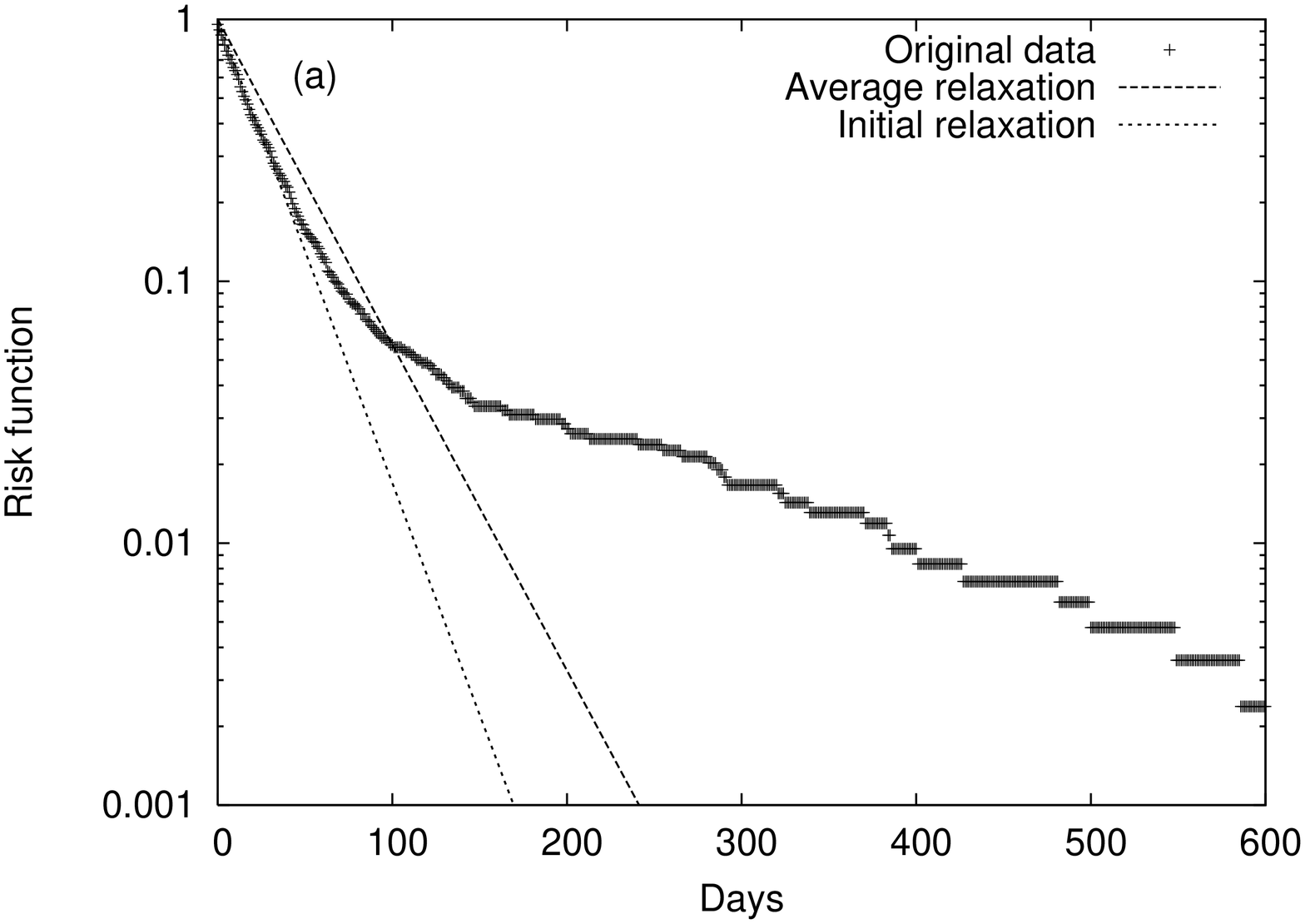}

\includegraphics[angle=-90,width=5.00in]{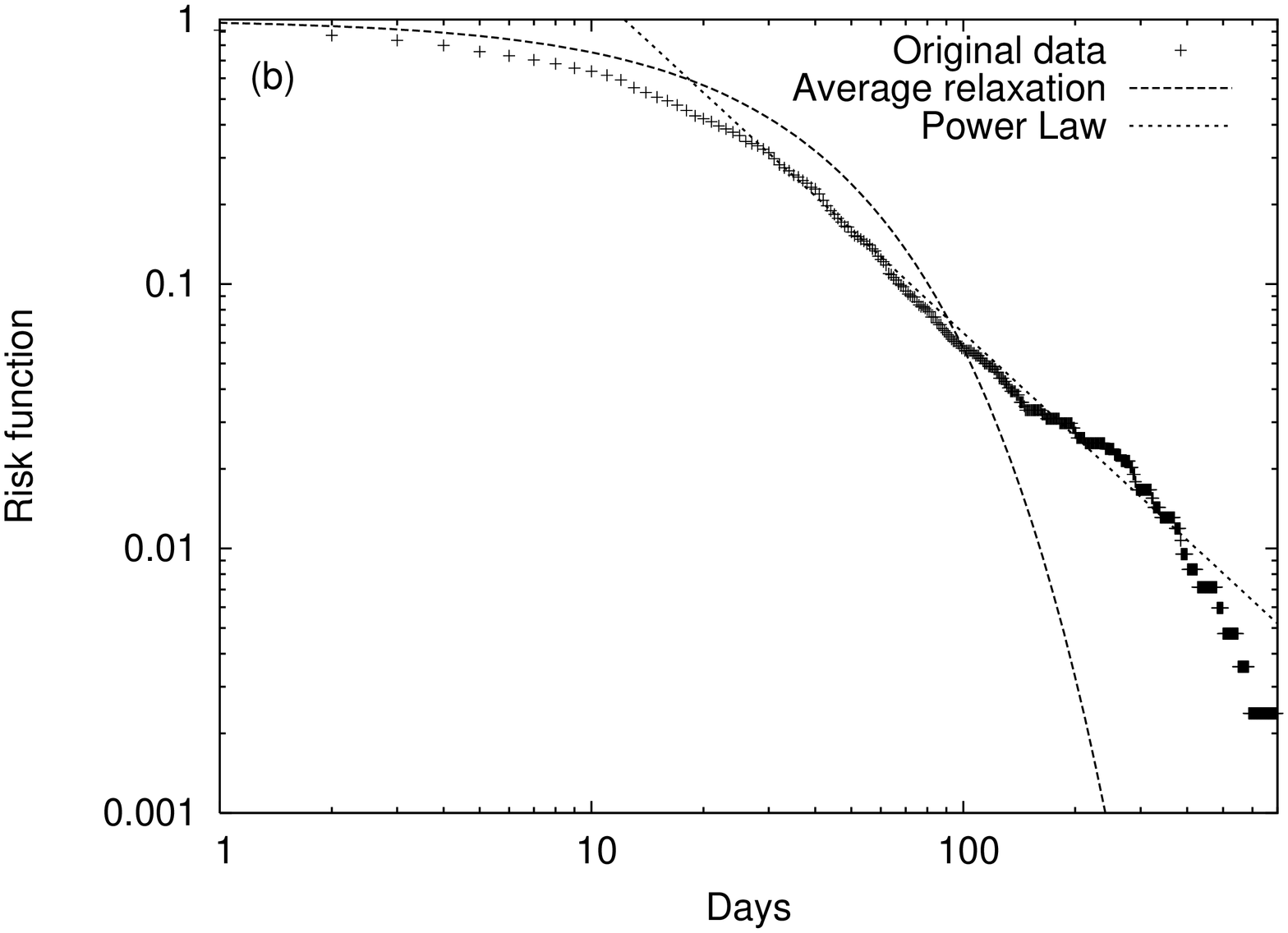}

\caption{\label{distributionT} Risk functions $G(t)$, on (a) semi-log  and (b) log-log scales.
  (a) The dashed lines represent the average relaxation $e^{-\lambda_A t}$, associated to the exponential  distribution $f(\tau) = \lambda_A e^{-\lambda_A t}$, and the initial exponential relaxation Eq.(\ref{init}). In (b) the power law regime, Eq.(\ref{medium}),  is compared to the average relaxation as in (a). }
\end{figure}

The empirical risk function is plotted in Fig.(\ref{distributionT}) together with the corresponding  exponential distribution $f(\tau) = \lambda_A \exp(-\lambda_A \tau)$,  on (a) semi-log, and (b)  log-log scales.

\begin{figure}

\includegraphics[angle=-90,width=5.00in]{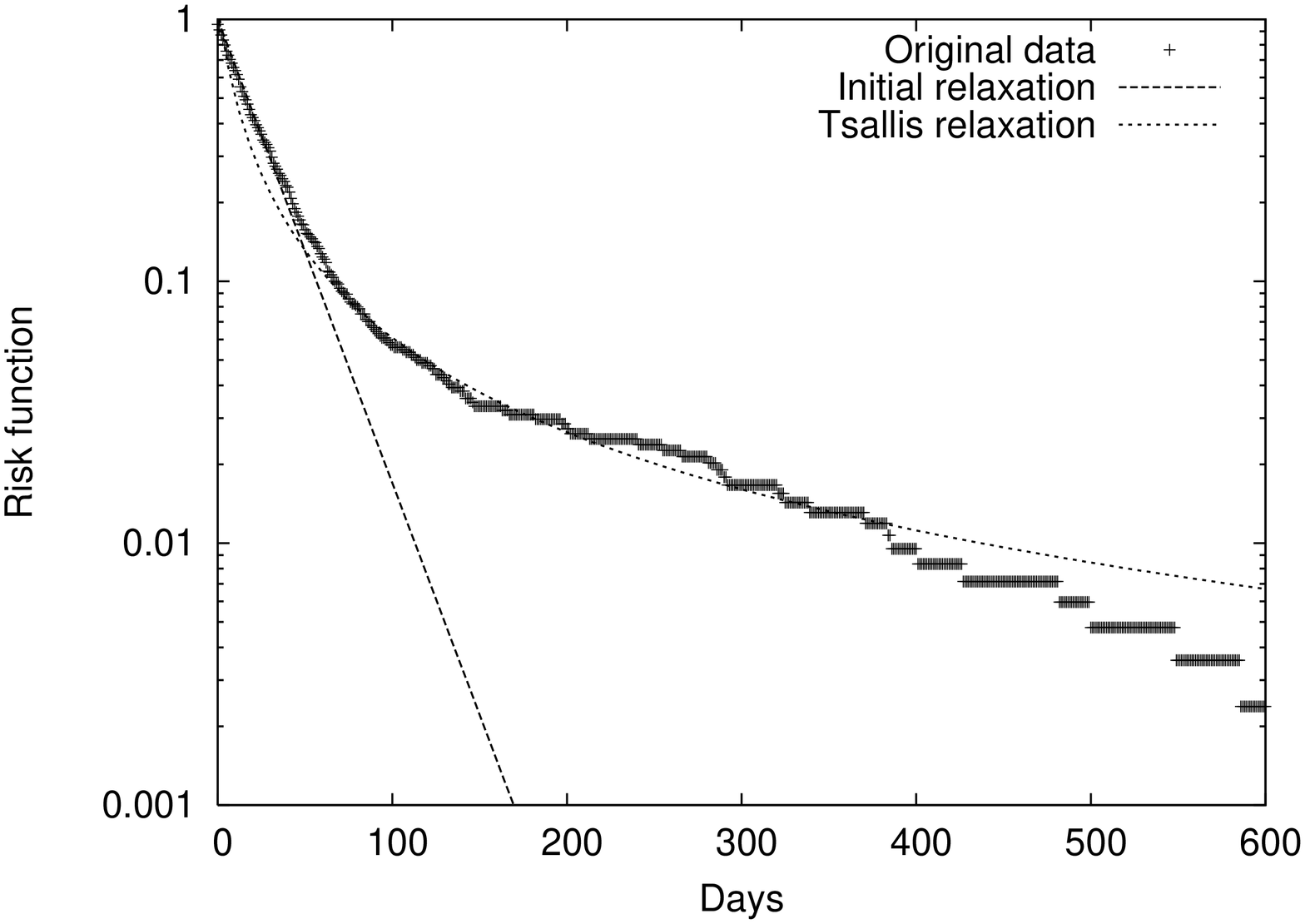}

\caption{\label{distributionTsallis} Risk function $G(t)$ in semi-log scale.
  The dash line represents the initial relaxation $e^{-\lambda_{init} t}$; the Tsallis relaxation Eq.(\ref{tsallis}) is represented by a dotted curve; the Tsallis exponent is $\alpha_m= 1.3$ and the delay time is $t_T=12$. }
\end{figure} 

Obviously, the collected data and the average  relaxation differ from each other. However, for small times, the relaxation appears to be an exponential. We obtain the relaxation coefficient through a fit over 15 days (Fig.(\ref{distributionT}(a)) with:
\begin{equation}
\label{init}
G(t) = e^{-\lambda_{init} t} ~, ~ \lambda_{init}=0.04~, ~ \tau_{init}=25 (days)
\end{equation}
where $\tau_{init} = 1/\lambda_{init}$ is the relaxation time for small time lags.
It is also important to note, in the log-log scale,  (Fig.(\ref{distributionT}(b)), the power-law regime in the interval $[30:300]$, that we fit with:
\begin{equation}
\label{medium}
G(t) \sim t^{-\alpha_{m}} ~, ~ \alpha_{m}=1.3 .
\end{equation}

Such a behaviour, i.e. exponential for small values and power-law for large values, is compatible with a Tsallis distribution for the waiting times $f_{Tsallis}(t) = t_T^{\alpha_m} \alpha_m  (t_T+ t)^{-(\alpha_m + 1)}$,
where $t_T$ is a positive parameter that smoothens the behaviour at short times, and $(\alpha_m + 1)$ is the exponent of the power-law tail of the distribution. Let us also stress that Tsallis distributions conserve their form at the level of the risk function: 
\begin{equation}
\label{tsallis}
G_{Tsallis}(t) = t_T^{\alpha_{m}} (t_T + t)^{-\alpha_{m}} .
\end{equation}

In figure (\ref{distributionTsallis}), we verify that Eq.(\ref{tsallis}) fits very well the data up to 350 days, with $t_T=12$ and $\alpha_m=1.3$.
For larger times $\tau>400$, in contrast, the power-law regime and the Tsallis relaxation behaviour clearly cease to be true: there is a rapid deceleration that we associate with an exponential truncation, - . There is no reason that a time interval would be infinite indeed.
This behaviour reminds us of the results obtained for the velocity distribution of  a Brownian particle with a fluctuating mass \cite{RLMAunpublished}; the mass of the particle being here analogous to the number of planes at a given time. 
As so observed we can describe the time interval statistics within such a formal idea, i.e. 
the system is characterized by a  distribution with a varying characteristic time $\tau_c$ or relaxation coefficient $\lambda_c$.

\begin{figure}

\hspace{2cm}
\includegraphics[width=4.00in]{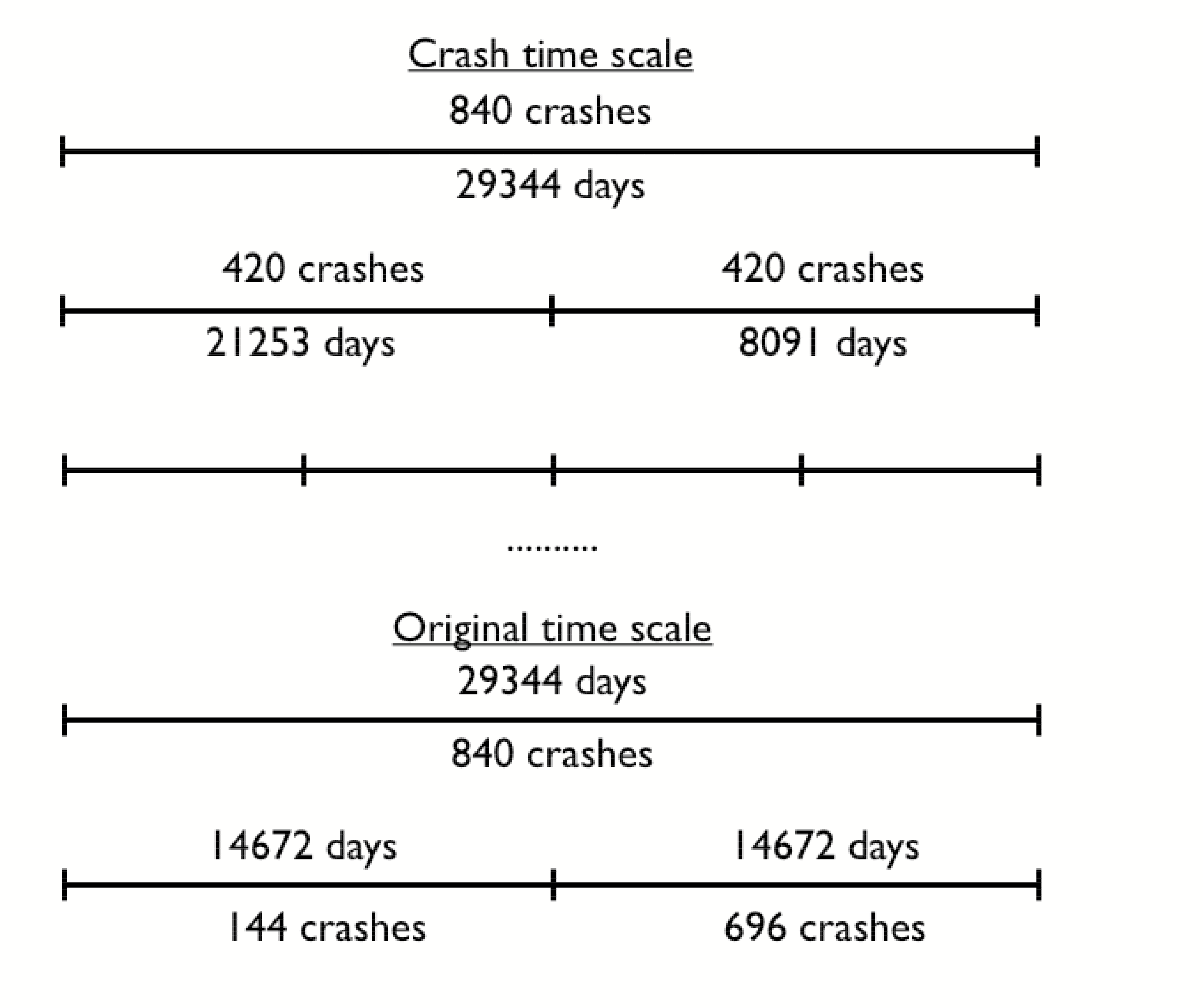}

\caption{\label{procedure} Procedure in order to extract the time dependence of $\tau_c$. In the crash time scale, we get $\tau_{c1}=50.6$ and $\tau_{c2}=19.26$, that verifies $\frac{1}{2} (\tau_{c1}+\tau_{c2})=\tau_A$ $\sim 35$. In the natural time scale, however, we get $\tau_{c1}=101.88$, $\tau_{c2}=21.08$ and 
$\frac{1}{2} (\tau_{c1}+\tau_{c2}) \neq \tau_A$.}
\end{figure} 

\begin{figure}
\includegraphics[angle=-90,width=5.00in]{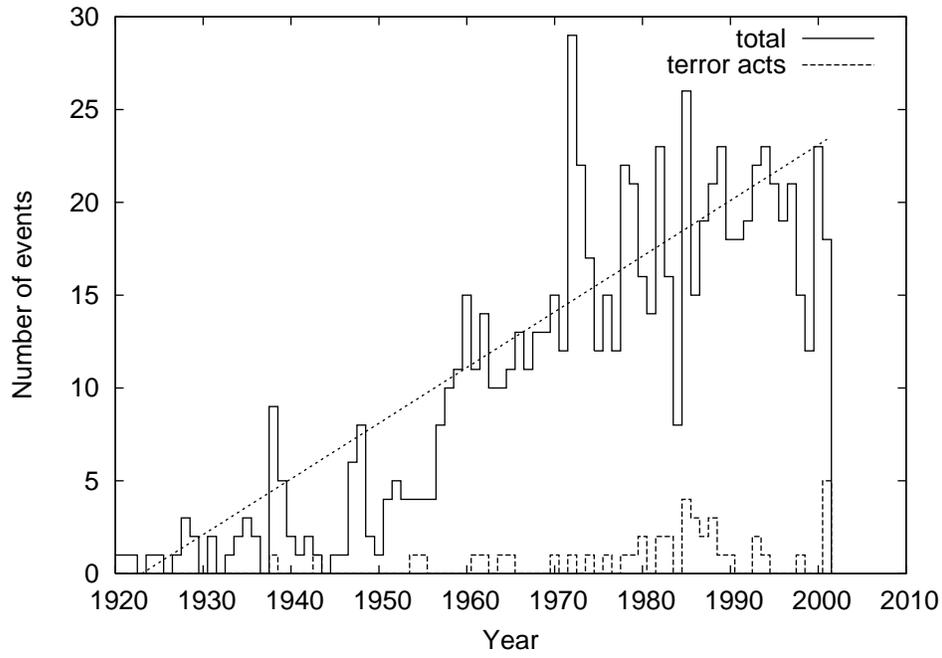}
\caption{\label{ausloos} Time evolution of the number of crashes per year since the beginning of commercial airline history.}
\end{figure} 

\begin{figure}
\includegraphics[angle=-90,width=5.00in]{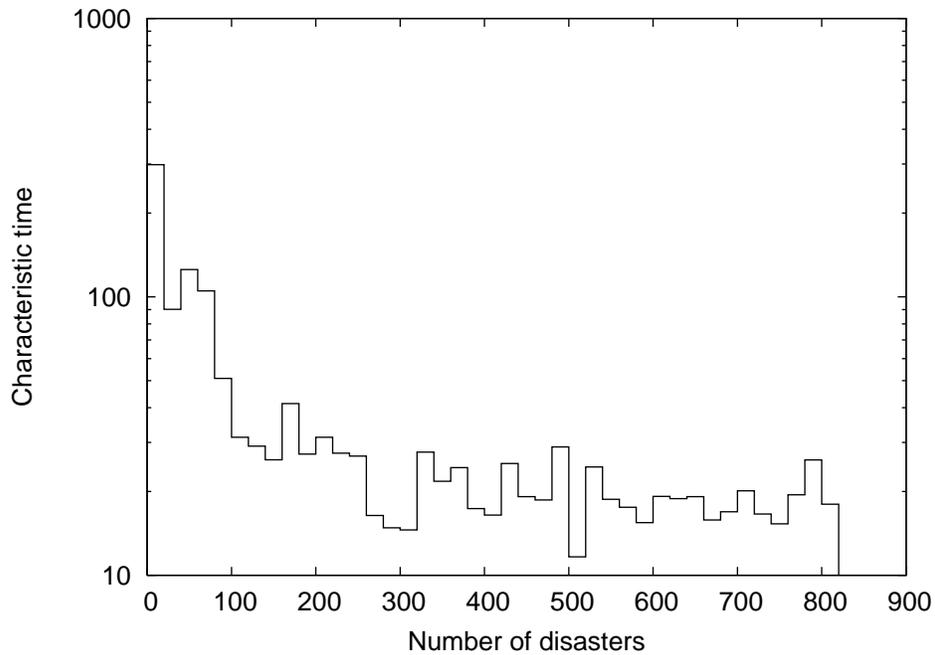}
\caption{\label{localT} Time evolution of the characteristic times $\tau_c$, where the time is counted in days through the number of airline disasters which have occurred since the beginning of commercial airline history.}
\end{figure} 

\begin{figure}

\includegraphics[angle=-90,width=5.00in]{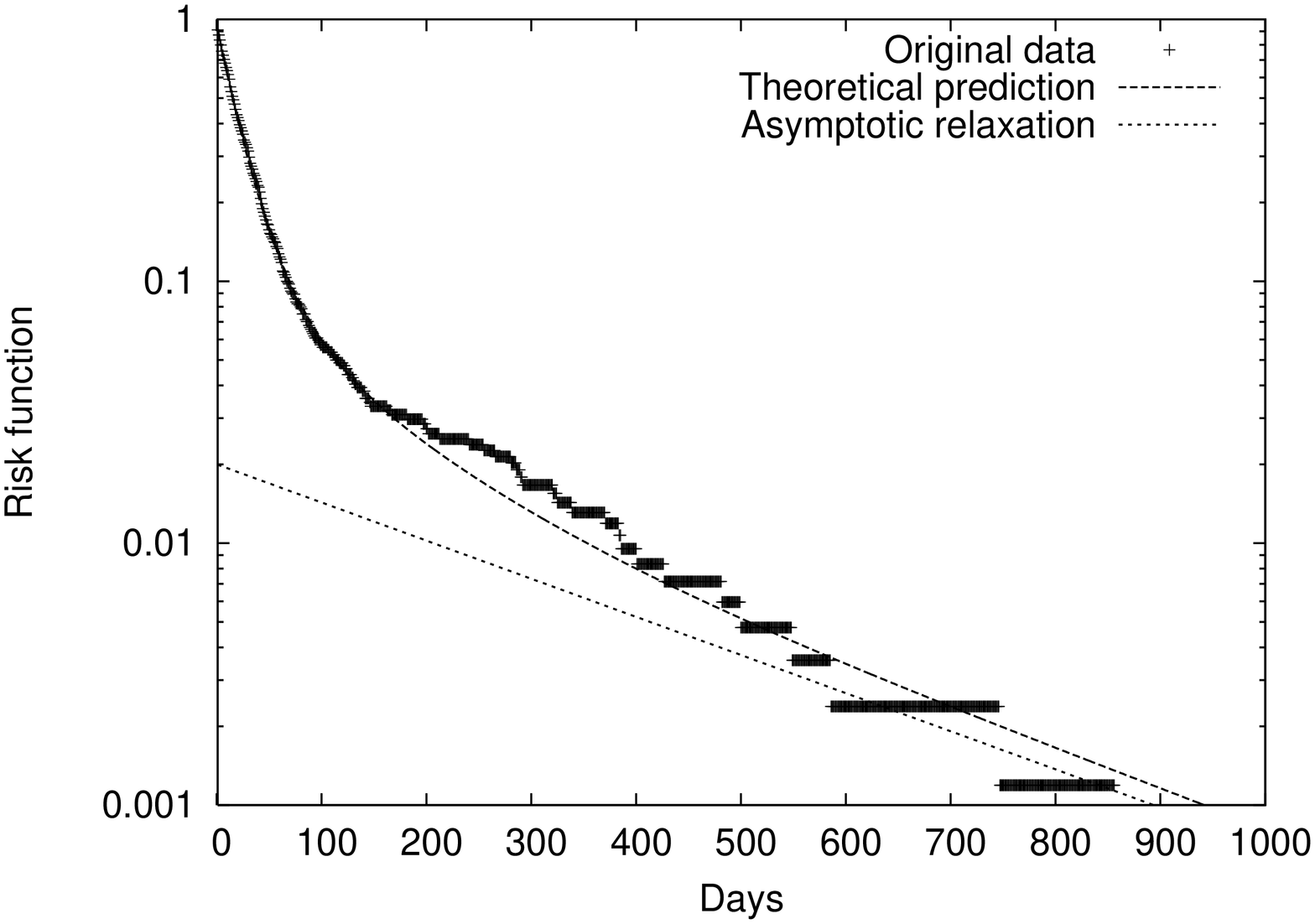}

\caption{\label{beck} Risk functions $G(t)$ on log-normal  plot.
  The dashed lines represent the theoretical prediction, Eq.(\ref{theo}) and the asymptotic relaxation, Eq.( \ref{asym}).}
\end{figure} 

In order to show the   time dependence of 
the varying characteristic time $\tau_c$, we apply the  procedure explained in Fig. \ref{procedure}. It consists in dividing the signal into windows, and to measure $\tau_c$ locally in time. Two options are possible: either divide the system into windows of equal time  intervals or  have time bins with an equal number of crashes. The choice corresponds to measuring the time in usual (or natural) units (days), or in ''number of disasters'' respectively. Let us stress that these two possible descriptions also take place in 
describing granular gases \cite{granular}, where one may measure time in seconds, or in number of collisions, depending on the purpose. Such a choice also occurs when discussing whether power laws exist in financial data  \cite{serva}. Here below, we prefer to apply the second approach, namely we count time in ''number of disasters'' $N_D$, and we divide the data in $K_i$ intervals encompassing the same number of collision $C$, each of the intervals being characterised by a {\em local} characteristic time $\tau_{ci}$.  

Recall that the analogous thermodynamic system is considered to be out-of-equilibrium with a non conserved sort of mass, i.e. the number of planes or flights. It can be easily noticed that the number of disasters increases with time, - roughly following a linear law (Fig. 6); a best least square fit  ($R \simeq 0.87$) gives a slope $\sim 0.3$ ($year^{-1}$). This is likely to be understood as being related to the steady increase in the number of flights and/or passengers. Indeed quoting \cite{wiki} ($http://en.wikipedia.org/wiki/Airline\#Development\_of
\_air${\em -} 
$lines\_post${\em-1945}) : ''The demand for air travel services is derived demand. ...  Notwithstanding (these) demand patterns, the overall trend of demand has been consistently increasing. In the 1950's and 1960's, annual growth rates of 0.15 or more were common. Annual growth of 0.05-0.06 persisted through the 1980's and 1990's.'' However we have not been able to find the true meaning of these growth rates. It would be interesting to know the true growth rate of passenger flights and compare it to the here above empirical slope value. This might point (or not) to recent considerations on ''black list'' or not of airline traffic and accidents.  Notice that a measure of deaths through airplane crash probability per flight is known to be 1/25000, according to standard insurance practice. However $http://www.planecrashinfo.com/cause.htm$ pretends that the odds of being killed in a single trip is 1/73187 if only data from  NTSB Accidents and Accident Rates by NTSB Classification  \cite{NTSB} between  1995 and 2004 is taken into account.

Finally, let us stress that the chosen procedure, i.e. measuring time in number of disasters, seems the best to take into account this ''growth trend'' in the data analysis.
Indeed, the method allows to preserve the average time lag between these (see Fig.{\ref{procedure}}):
\begin{equation}
\label{scale}
\frac{1}{K_i} \sum_{i}^{K_i} \tau_{ci}  = \tau_A
\end{equation}
 where the sum is performed over all intervals $K_i$. This property is  not verified by using the natural time scale. For a lengthier  discussion on the definition of time,  calendar time or business time 
 (in foreign exchange markets), in view of modeling the dynamics and searching for (scaling) laws see \cite{serva}.
 
 In figure \ref{localT}, we plot the time evolution of $\tau_c$, by dividing the system into windows of 20 disasters. Empirical results clearly exhibit the time dependence of $\tau_c$, i.e. $\tau_c$ has decreased by a factor 10 over the last century. This confirms the increase of traffic, $\sim$ {\em increase of mass}, during that time period.

Let us now assume that the waiting times in each window are exponentially distributed 
\begin{equation}
f_i(\tau)= \tau_{ci}^{-1} \exp{(- \tau/ \tau_{ci}).}
\end{equation}
Consequently,  the average distribution for waiting times is 
\begin{equation}
\label{theo}
\overline{f}(\tau)= \frac{1}{K_i} \sum_{i}^{K_i} \tau_{ci}^{-1} 
 \exp{(- \tau/\tau_{ci}).}
\end{equation}
In the long time limit, this distribution is controlled by the largest value of $\tau_{ci}$. In our example, i.e. division into windows of 20 disasters, it is $\tau_{c1}=298$ days$^{-1}$. This means that the tail of the distribution is made of events occurring in the beginning of the XX century, when airline disasters were ... rare events 
\begin{equation}
\label{asym}
\overline{f}(\tau) \rightarrow  \exp{(- \tau/\tau_{c1})}
\end{equation}
Indeed, at the beginning of the XX century there was one crash per $year$ approximatively, while there is approximately two such disastrous  events per $month$ nowadays (see Fig.6) 

We verify that these formulae describe successfully the empirical data in figure \ref{beck}, thereby showing that airline disasters are well-described by a time dependent  process. The success of the method also enlightens the important role of the non-stationarity of a random variable in the emergence of anomalous distributions, and gives a direct application to the abstract formalism developed in \cite{RLMAunpublished}.
One should also note that this formalism provides an elegant explanation to the deviations to the Poissonian observed in Fig.2. Indeed, deviations for small values of $n$ are due to the very rare events of the beginning of last century, that enhance the importance of low values of $n$ in the empirical distribution. 

\section{Terrorism vs. Other Origin}

\begin{figure}

\includegraphics[angle=-90,width=5.00in]{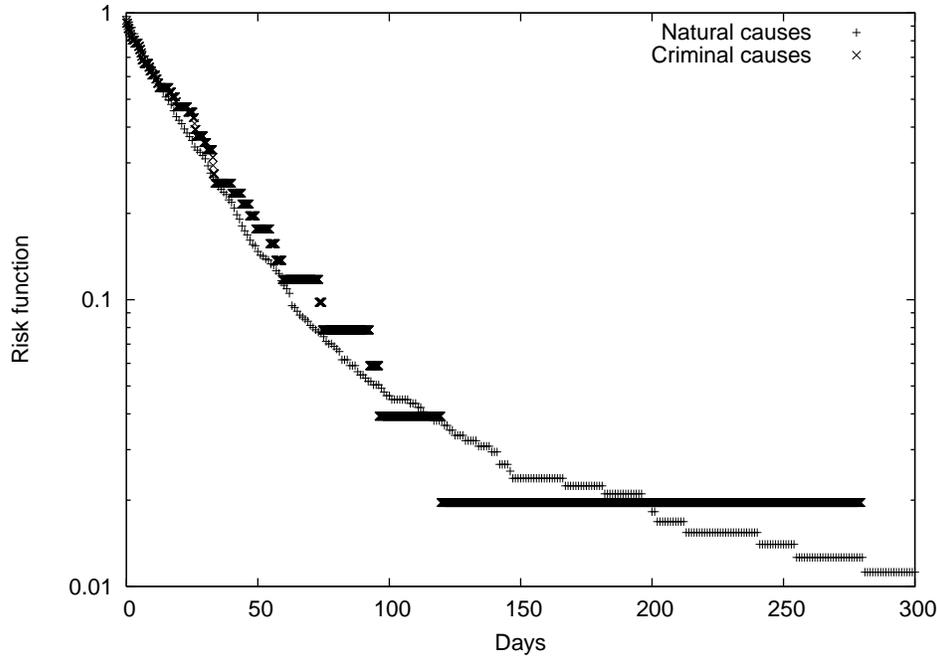}

\caption{\label{beck2} Risk functions $G(t)$ in log-normal  scale, for criminal and natural causes; see text for scaling argument.}
\end{figure} 

In order to examine  some sort of external field effect, i.e. criminal (sometimes called  terrorist) acts on plane accident time lag distribution, we
  divide the events into two categories: those considered to have been occurring because of ''natural'' reasons (geophysical cause, human error, ...), and those due to criminal reasons (so called terrorist attack, hijacking, ...).  The selection is made according to the comments on ''disaster cause column'' in the table of the studied website data. Again some caveat is in order since it is not always proved or admitted that there is a terrorist act associated with a plane loss. Neither is it usually admitted that there is a ''military" mistake, like shooting a missile to a commercial civilian plane, either accidentally or voluntarily (because the plane flight passes over a military zone). The ''sabotage cause'' is considered to be 8 per cent by  $http://www.planecrashinfo.com/cause.htm$.
  
 By using the comments in the data table, and admitting that the data might contain some omission or incorrect information,  we found  52 {\em criminal crashes} and 789 {\em natural crashes},  
  so that their average time intervals are 564 days and 37 days respectively.  The time sequence of the disasters due to  such acts is shown in Fig. 5. In order to  compare the time lag distributions for these two kinds of events, we have calculated the two corresponding risk functions. Moreover, we have rescaled their time scale so that their distributions have the same average $\tau_A$. Practically,  we rescale the time scale of {\em criminal crashes} by a factor $37/564$. The resulting risk functions are shown in Fig. 9.
  Given the precision of the data, we observe that the time distributions look the same in the cases of ''natural causes' and so called 'terrorist acts''. This suggests that the same time dependent exponential  distribution can also be considered as a best fit to both data.

\section{Conclusions}

''An aviation accident (as per the U.S. National Transportation Safety Board definition) is an occurrence associated with the operation of an aircraft which takes place between the time any person boards the aircraft with the intention of flight and all such persons have disembarked, and in which any person suffers death or serious injury, or in which the aircraft receives substantial damage, while an aviation incident is an occurrence other than an accident, associated with the operation of an aircraft, which affects or could affect the safety of operations. Other countries adopt a similar approach, although there are minor variations, such as to the extent of aviation-related operations on the ground covered, as well as with respect to the thresholds beyond which an injury is considered serious or the damage is considered substantial'' \cite{wiki}.

We have examined the time lag distribution between  airplane flown by commercial airline disasters
 as recorded on a freely available web site. We have warned about the validity of the data, sometimes due to conditions out of the responsability of the webmaster.  We have examined whether the distributions
  differ when so called terrorist (more generally criminal) acts are taken into account. There seems to be no influence of any violent act
   on the total distribution. We suggest that the distribution be described as a time dependent   exponential distribution function.
    A connection with a microscopic statistical physics framework can be found through the Tsallis distribution function,
     - since we can consider that the number of airplanes is like the time dependent mass of a particle undergoing a Brownian motion.

There is no truly ''microscopic" modeling at this time, in view of the lack of information on possible causes,
 though it might be interesting to correlate the crash or disaster to e.g. pilot biorythms, sun spot frequencies, etc.,
  as done in other cases, like car accidents \cite{MARALD}. 
  Other ''internal degrees of freedom'' can be used to distinguish cases.
   The ''volume'' of casualties could also be studied.
   
   Notice that a time dependent Poisson distribution has been found in many cases : epidemics  \cite{Reilly}, 
   social studies \cite{Holden}, finance \cite{Takayasu}, and even in the air  transportation industry \cite{Hebert} but also for water distribution
  \cite{Buchberger}, 
   for coronal mass ejection \cite{Moon2002} and
   for solar flares waiting times \cite{originsolarflares}, with quite different time scales of course. However it is known that
   the density of solar flares is roughly characterized by an eleven year cycle. The recent maxima occurred in 1969, 1980, 1991 and 2002
   \cite{sunspothistory}.
   Observing that 1972 is the year with the maximum number
   of disasters for the examined data, a rapid calculation indicates that the years 1961, 1982, 1994 and 2005 should be close
   to a maximum in the number of crash density if an 11-year cycle is postulated. This is the case indeed. 
   Recall also that the sun has a 27 day rotation period.

(Comment inserted after analysis completion:
During the last months that this report was waiting to be prepared for publication, several disasters occurred, i.e. 
Toronto, Palermo, Athens, Machiques, Pucallpa, Singapore, ... apparently without   ist acts. Some follow up of the above work seems of interest.)

{\bf Acknowledgements}

Part of RL work 
has been supported by European Commission Project 
CREEN FP6-2003-NEST-Path-012864.
 Part of this work results from  financing through the ARC 02-07/293 Project of the
 ULg 
  which MA also thoroughly acknowledges. Critical comments by D. Stauffer have been as always
  very valuable for improving this report.


\begin{thebibliography}{0}


\bibitem{gregory} P. Gregory, Bayesian Logical data Analysis for the Physical Sciences, Cambridge U.P., Cambridge, 2005.

\bibitem{poissondistrwebsite} $http://www.gfi.uib.no/$$\sim$$nilsg/kurs/notes/node32.html$

\bibitem{wiki} $http://en.wikipedia.org/wiki/Airline\#Development\_of\_airlines\_post-1945$ 
 
 \bibitem{NTSB} $http://www.planecrashinfo.com/cause.htm$
 
\bibitem{lauda}   $http://www.mahk.com/sc987.htm$ 

\bibitem{RLMAPhysAshocks} R. Lambiotte, M. Ausloos, Physica A, in press

\bibitem{sornetteofcourse} D. Sornette, F. Deschatres, T. Gilbert, Y. Ageon  Phys. Rev. Lett. 
  93  (2004) 228701 

\bibitem{Tsallisdistribution} C. Tsallis, J. Stat. Phys. 52 (1988) 479 

\bibitem{tsallis2} C. Tsallis, Braz. J.  Phys. 29 (1999) 1

\bibitem{datawebsite} {\em http://dnausers.d-n-a.net/dnetGOjg/Disasters.htm}  

\bibitem{RLMAunpublished} R. Lambiotte, M. Ausloos,  arxiv cond-mat/0508773

 \bibitem{siteipa}  {\em http://www.infoplease.com/ipa/A0001449.html}

 \bibitem{sitescaruffi} {\em http://www.scaruffi.com/politics/aircrash.html}.

\bibitem{granular} R. Soto, M. Mareschal, Phys. Rev. E 65 (2001) 041303 

 \bibitem{serva}  L. Berardi, M. Serva, Physica A 353 (2005) 403
 
\bibitem{MARALD} M. Ausloos, R. Ausloos, L. Demarthe, unpublished.

\bibitem{Reilly} C. Reilly, T. Schacker, A.T. Haase, S. Wietgrefe, D. Krason, 
  J. Amer. Stat. Ass. 97 (2002) 943

\bibitem{Holden}  R.T. Holden,   Sociological Methods and Research 14  (1985) 3

\bibitem{Takayasu} M. Takayasu,   H.Takayasu,  Physica A 324 (2003) 101

\bibitem{Hebert} J.E. Hebert,  D.C. Dietz,  J. Aircraft 34 (1997) 43

\bibitem{Buchberger} S.G. Buchberger,  G. J. Wells, J. Water Res. Plan. Manag. 122 (1996) 11

\bibitem{Moon2002} Y.J. Moon, G.S. Choe, H.  Wang,  Y. D.  Park, Astrophys. J. 581 (2003) 1176

\bibitem{originsolarflares} M. S. Wheatland, Astrophys. J. 536 (2000) L109

\bibitem{sunspothistory} $http://www.spaceweather.com/java/sunspot.html$

\end{thebibliography}
\end{document}